\documentclass[12pt,a4paper]{article}%
\usepackage{latexsym}
\usepackage{amsmath}
\usepackage{amsfonts}
\usepackage{amssymb}
\usepackage{graphicx}%
\providecommand{\U}[1]{\protect\rule{.1in}{.1in}}
\begin{document}

\title{Vortices as Instantons in Noncommutative Discrete Space: Use of $Z_{2}$ Coordinates}
\author{\textsf{Hideharu Otsu } \thanks{otsu@vega.aichi-u.ac.jp}\\Faculty of Economics, Aichi University,\\Toyohashi, Aichi 441-8522, Japan
\and \textsf{Toshiro Sato} \thanks{tsato@mie-chukyo-u.ac.jp}\\Faculty of Law and Economics, Mie Chukyo University, \\Matsusaka, Mie 515-8511, Japan
\and \textsf{Hitoshi Ikemori } \thanks{ikemori@biwako.shiga-u.ac.jp}\\Faculty of Economics, Shiga University, \\Hikone, Shiga 522-8522, Japan
\and \textsf{Shinsaku Kitakado }\thanks{kitakado@ccmfs.meijo-u.ac.jp}\\Department of Physics, Faculty of Science and Technology,\\Meijo University,\\Tempaku, Nagoya 468-8502, Japan}
\date{}
\maketitle

\begin{abstract}
We show that vortices of Yang-Mills-Higgs model in $R^{2}$ space can be
regarded as instantons of Yang-Mills model in $R^{2}\times Z_{2}$ space. For
this, we construct the noncommutative $Z_{2}$ space by explicitly fixing the
$Z_{2}$ coordinates and then show, by using the $Z_{2}$ coordinates, that BPS
equation for the vortices can be considered as a self-dual equation. We also
propose the possibility to rewrite the BPS equations for vortices as ADHM
equations through the use of self-dual equation.

\end{abstract}

\newpage

\section{Introduction}

Topological solitons play an important role in various field theories. These
are kink, vortex, baby-skyrmion, monopole, skyrmion, instanton and so on
\cite{Actor}. Some of the soliton equations are solved analytically, others
are solved only numerically. It is interesting to look for the relations among
the topological solitons. We consider a static soliton in Yang-Mills-Higgs
(YMH) model in $2+1$ dimensions. The static soliton is a vortex in
2-dimensional $R^{2}$\ space. Some properties of Abelian vortex and
non-Abelian vortex in YMH model have been studied
\cite{Nielsen:1973cs,Weinberg,Tong}. The vortex configurations are solved
numerically. The BPS (Bogomol'nyi-Prasad-Sommerfield) equations\ \cite{BPS}
for the vortex can be rewritten in terms of master equation plus
half-ADHM(Atiyah-Drinfeld-Hitchin-Manin) equation \cite{Eto}. The solution of
half-ADHM equation contains information on the moduli space of the vortex,
while instanton in 4-dimensional space are solved analytically by the ADHM
method \cite{Atiyah:1978ri}.

On the other hand,\textit{ }Higgs fields can be treated as gauge fields
\cite{Coquereaux}. Note that, in these works discrete spaces are treated in
terms of differential forms without the explicit use of the
coordinates.\textit{ }We have been investigating the possibility of describing
a vortex in 2-dimensional space as an instanton in 4-dimensional space, which
is $R^{2}\times Z_{2}$ space\ in this paper. In the previous paper
\cite{Ikemori:2008pb}, from a viewpoint of the noncommutative differential
geometry and gauge theory in discrete space, we have shown that the instanton
in $R^{2}\times Z_{2}$ space is nothing but the vortex in $R^{2}$ space. This
means that difference of vortex and instanton can be considered as that of the
spaces $R^{2}\times Z_{2}$ and $R^{4}$. In ref. \cite{Ikemori:2008pb}, we did
not explicitly discuss the relations between the Yang eq. and the master eq.,
due to lack of representation of the $Z_{2}$ coordinates. The ADHM method for
vortices also requires the coordinate representation. By introducing the
explicit form of the $Z_{2}$ coordinates, we can approach the problem from the
new point of view. An attempt of this paper is the analysis using the explicit
form of the noncommutative $Z_{2}$ coordinates. On the other hand, the
arguments with the differential forms can not be cast straightforwardly into
the coordinate picture. The purpose of this paper is to clarify the relation
between instanton and vortex using\ the noncommutative $Z_{2}$\ coordinates.
We first define the coordinates for noncommutative $Z_{2}$\ space and then
investigate the relation between the instantons in $R^{2}\times Z_{2}$\ space
and the vortices in $R^{2}$ space. In addition, we consider the relations
among different descriptions of the vortices.

In section 2, we summarize properties of YMH model and fix the notations. In
section 3, we construct a noncommutative $Z_{2}$ space. In section 4, we
discuss relation between the instanton in $R^{2}\times Z_{2}$ space and the
vortex in YMH model in $R^{2}$ space. In section 5, we discuss relations among
BPS, master and half-ADHM equations. The final section is devoted to summary
and discussion.

\section{Some properties of Yang Mills Higgs model}

Let us summarize here some properties of the YMH model which has non-Abelian
gauge symmetry \cite{Ikemori:2008pb}. The model contains a Higgs field,
represented by $N_{L}\times N_{R}$ matrix, and two gauge fields corresponding
to $U\left(  N_{L}\right)  \times U\left(  N_{R}\right)  $ gauge group. In
this paper, we consider the models with $N_{L}=N_{R}=N$, where the solitons
are local vortices. The Lagrangian in $2+1$ dimensions is
\begin{align}
{\mathcal{L}}  &  =\mathbf{Tr}\left(  \frac{1}{2g^{2}}\left(  F^{L}\right)
^{\mu\nu}\left(  F^{L}\right)  _{\mu\nu}+\frac{1}{2g^{2}}\left(  F^{R}\right)
^{\mu\nu}\left(  F^{R}\right)  _{\mu\nu}\right) \nonumber\\
&  +\mathbf{Tr}\left(  \left(  D_{\mu}H\right)  ^{\dag}D^{\mu}H-\frac{g^{2}%
}{2}(HH^{\dag}-c\boldsymbol{1}_{N})^{2}\,\right)  . \label{YMH-L}%
\end{align}
Where, we define a covariant derivative $D_{\mu}$ and field strength
$F_{\mu\nu}^{L}$ , $F_{\mu\nu}^{R}$\ as%
\begin{equation}
D_{\mu}H=\partial_{\mu}H+L_{\mu}H-HR_{\mu}\;,
\end{equation}%
\begin{align}
\left(  F^{L}\right)  _{\mu\nu}  &  =\partial_{\mu}L_{\nu}-\partial_{\nu
}L_{\mu}+\left[  L_{\mu},L_{\nu}\right]  \ ,\\
\left(  F^{R}\right)  _{\mu\nu}  &  =\partial_{\mu}R_{\nu}-\partial_{\nu
}R_{\mu}+\left[  R_{\mu},R_{\nu}\right]  \;,
\end{align}
and $\mathbf{Tr}$ is a trace over the adjoint representation of $U\left(
N\right)  .$ Two $U\left(  N\right)  $ gauge fields $L_{\mu}$ , $R_{\mu}$\ and
the Higgs field $H$ are represented by $N\times N$ matrices. In the following
we take $g^{2}=2$ and $c=1$ for simplicity. The energy integral is of the
form
\begin{equation}
E=\int dx_{1}dx_{2}\mathbf{Tr}\left(  \frac{1}{2}\left\vert F_{12}%
^{L}\right\vert ^{2}+\frac{1}{2}\left\vert F_{12}^{R}\right\vert
^{2}+\left\vert D_{1}H\right\vert ^{2}+\left\vert D_{2}H\right\vert
^{2}+\left(  HH^{\dag}-\boldsymbol{1}_{N}\right)  ^{2}\right)  . \label{YMH-E}%
\end{equation}
The BPS equations minimizing the energy are
\begin{align}
\left(  D_{1}\pm iD_{2}\right)  H  &  =0\ ,\label{BPS1}\\
iF_{12}^{L}\pm\left(  HH^{\dag}-\boldsymbol{1}_{N}\right)   &
=0\ ,\label{BPS2}\\
iF_{12}^{R}\mp\left(  H^{\dag}H-\boldsymbol{1}_{N}\right)   &  =0\;,
\label{BPS3}%
\end{align}
where we use the anti-Hermitian gauge fields $L_{\mu}^{\dagger}=-L_{\mu}$ and
$R_{\mu}^{\dagger}=-R_{\mu}$ [4]. The solutions of the equations (\ref{BPS1}),
(\ref{BPS2}), (\ref{BPS3}) are topologically stable solitons, called
non-Abelian vortices. Where, 2 sets of equations are those for vortex and for
anti-vortex. It is obvious that only pure gauge configurations are allowed at
the spacial infinity $\left\vert x\right\vert \rightarrow\infty$. This means
that the topological property of the non-Abelian vortices is classified by the
mapping index for $S^{1}\rightarrow$ $U\left(  N\right)  \times U\left(
N\right)  .$ On account of the fact that $U\left(  N\right)  \ $is equal to
$U\left(  1\right)  \times SU\left(  N\right)  $, the corresponding homotopy
group is
\begin{equation}
\pi_{1}\left(  U\left(  N\right)  \times U\left(  N\right)  \right)  =\pi
_{1}\left(  U\left(  1\right)  \times U\left(  1\right)  \right)
=\mathbb{Z\times Z\;}. \label{homotopy}%
\end{equation}
We can take the topological charges corresponding to (\ref{homotopy}) as%
\begin{equation}
Q_{L-R}\equiv\dfrac{i}{2\pi}\int dx_{1}dx_{2}\mathbf{Tr}\left(  F_{12}%
^{L}-F_{12}^{R}\right)  =0,\pm1,\pm2,\cdots\label{Qv}%
\end{equation}
and%
\begin{equation}
Q_{L+R}\equiv\dfrac{i}{2\pi}\int dx_{1}dx_{2}\mathbf{Tr}\left(  F_{12}%
^{L}+F_{12}^{R}\right)  =0,\pm1,\pm2,\cdots\ .
\end{equation}
Here, $Q_{L-R}$ is identified with the vortex number. On the other hand,
topological charge $Q_{L+R}$ is irrelevant to the vortex configuration, since
gauge field \textrm{$\mathbf{Tr}$}$\left(  L_{\mu}+R_{\mu}\right)  $ does not
interact with other fields. Although the general configurations are classified
by two topological charges $Q_{L-R}$ and $Q_{L+R}$, the vortex configurations
are essentially classified by $Q_{L-R}$. Because the BPS equations
(\ref{BPS2}) and (\ref{BPS3}) mean that the $U(1)$ part of $F_{12}^{L}+$
$F_{12}^{R}=0$, and thus $Q_{L+R}=0$ for the vortex solutions.

Note that, although our YMH models have the $U\left(  N\right)  \times
U\left(  N\right)  $ gauge group with $1\leq N$, vortex solutions have some
relations to those of the model with $U\left(  N\right)  $ gauge group.
Particularly, the model of $U\left(  1\right)  _{L}\times U\left(  1\right)
_{R}$ gauge group is equivalent to the model of $U\left(  1\right)  _{L-R}$
gauge group, since one of the combinations of gauge field, i.e. $L+R$,
decouples from other fields. For $2\leq N$, the relations among vortex
solutions of the models with $U\left(  N\right)  $ and $U\left(  N\right)
_{L}\times U\left(  N\right)  _{R}$\ gauge groups are shown in section 5.

Let us describe the notations for 4-dimensional space, since we construct the
vortex in 2-dimensional space from a model in 4-dimensional space. The
relation between Cartesian coordinates $\left(  x_{1},x_{2},x_{3}%
,x_{4}\right)  $\ and complex coordinates $\left(  z,\bar{z},w,\bar{w}\right)
$ in 4-dimentional space are%

\begin{align}
z  &  =\frac{1}{\sqrt{2}}\left(  x_{1}+ix_{2}\right)  ,\;\bar{z}=\frac
{1}{\sqrt{2}}(x_{1}-ix_{2}),\nonumber\\
w  &  =\frac{1}{\sqrt{2}}\left(  x_{3}+ix_{4}\right)  ,\;\bar{w}=\frac
{1}{\sqrt{2}}(x_{3}-ix_{4}),\nonumber\\
\partial_{z}  &  =\frac{1}{\sqrt{2}}\left(  \partial_{1}-i\partial_{2}\right)
,\;\partial_{\bar{z}}=\frac{1}{\sqrt{2}}\left(  \partial_{1}+i\partial
_{2}\right)  ,\nonumber\\
\partial_{w}  &  =\frac{1}{\sqrt{2}}\left(  \partial_{3}-i\partial_{4}\right)
,\;\partial_{\bar{w}}=\frac{1}{\sqrt{2}}\left(  \partial_{3}+i\partial
_{4}\right)  . \label{coordinate}%
\end{align}
For $R^{4}$\ space, $\left(  z,\bar{z},w,\bar{w}\right)  $ are usual complex
coordinates. While, for $R^{2}\times Z_{2}$ space used in this paper, $w$ and
$\bar{w}$ are noncommutative coordinates to be defined in the next section.
Gauge fields are defined by
\begin{align}
a_{z}  &  =\frac{1}{\sqrt{2}}\left(  a_{1}-ia_{2}\right)  ,\;a_{\bar{z}%
}=-\frac{1}{\sqrt{2}}\left(  a_{1}+ia_{2}\right)  ,\label{fields}\\
\;a_{w}  &  =\frac{1}{\sqrt{2}}\left(  a_{3}-ia_{4}\right)  ,\;a_{\bar{w}%
}=-\frac{1}{\sqrt{2}}\left(  a_{3}+ia_{4}\right)  .
\end{align}
Finally, relations between the gauge field strength in the complex and
Cartesian coordinates are%
\begin{align}
F_{z\bar{z}}  &  =-iF_{12},\;F_{w\bar{w}}=-iF_{34},\nonumber\\
F_{z\bar{w}}  &  =-\frac{1}{2}\left(  F_{13}+iF_{14}+F_{24}-iF_{23}\right)
,\nonumber\\
F_{\bar{z}w}  &  =-\frac{1}{2}\left(  F_{13}-iF_{14}+F_{24}+iF_{23}\right)
,\nonumber\\
F_{zw}  &  =\frac{1}{2}\left(  F_{13}-iF_{14}-F_{24}-iF_{23}\right)
,\nonumber\\
F_{\bar{z}\bar{w}}  &  =\frac{1}{2}\left(  F_{13}+iF_{14}-F_{24}%
+iF_{23}\right)  . \label{leznov1}%
\end{align}

\section{Noncommutative $Z_{2}$ space}

In this section, we construct a 2-dimensional noncommutative discrete $Z_{2}$
space, referring to the construction of noncommutative $R_{NC}^{2}\ $\ space.
In the case of $R_{NC}^{2}\ $\ space, the complex coordinates\ are represented
by the creation and annihilation operators on the Fock space $\left\{
\left\vert n\right\rangle \right\}  $ with $n=0,1,2,3,\cdot\cdot\cdot$
\cite{Harvey}. Then the commutation relation of the complex coordinates is
proportional to the noncommutative parameter.

Now, we consider the coordinates $w$ and$\;\bar{w}$\ of noncommutative
discrete $Z_{2}$\ space as the operators on the Fock space with 2-states
$\left\vert 0\right\rangle $ and $\left\vert 1\right\rangle $. Our definition
of $Z_{2}$ space is%
\begin{equation}
w\left\vert 0\right\rangle =0,\;\bar{w}\left\vert 0\right\rangle =\sqrt
{\theta}\left\vert 1\right\rangle ,\;w\left\vert 1\right\rangle =\sqrt{\theta
}\left\vert 0\right\rangle ,\;\bar{w}\left\vert 1\right\rangle =0\;,
\label{z2}%
\end{equation}%
\begin{equation}
\bar{w}w\left\vert n\right\rangle =n\theta\left\vert n\right\rangle ,\;n=0,1,
\label{z22}%
\end{equation}
where $\theta$ is the noncommutative parameter. Then $Z_{2}$\ coordinates can
be represented by $2\times2$ matrices as%
\begin{equation}
w=\sqrt{\theta}\left(
\begin{array}
[c]{cc}%
0 & 1\\
0 & 0
\end{array}
\right)  ,\;\bar{w}=\sqrt{\theta}\left(
\begin{array}
[c]{cc}%
0 & 0\\
1 & 0
\end{array}
\right)  , \label{zzb-matrix}%
\end{equation}
where the Fock space is described by the vectors
\begin{equation}
\left\vert 0\right\rangle =\left(
\begin{array}
[c]{c}%
1\\
0
\end{array}
\right)  ,\;\left\vert 1\right\rangle =\left(
\begin{array}
[c]{c}%
0\\
1
\end{array}
\right)  .
\end{equation}
From (\ref{zzb-matrix}), coordinates $w$ and $\bar{w}$ are characterized by
anti-commutation relations%
\begin{equation}
\left\{  w,w\right\}  =\left\{  \bar{w},\bar{w}\right\}  =0 \label{N-C0}%
\end{equation}
and%
\begin{equation}
\left\{  w,\bar{w}\right\}  =\theta, \label{N-C}%
\end{equation}
where $\left\{  A,B\right\}  \equiv AB+BA$. Note that, the noncommutative
coordinates of $Z_{2}$ space satisfy the anti-commutation relations
(\ref{N-C0}) and (\ref{N-C}), in contrast to the case of $R_{NC}^{2}\ $space,
where the commutation relations $\left[  w,w\right]  =\left[  \bar{w},\bar
{w}\right]  =0$ and$\;\left[  \bar{w},w\right]  =\theta$ are satisfied. It
means that the coordinates of $Z_{2}$ space are fermionic, while those of
$R_{NC}^{2}$ space are bosonic.

Next, we define the differentiation by $w$ and$\;\bar{w}$\ as
\textquotedblleft right-differential\textquotedblright, namely differentiation
of a function $f\left(  w,\bar{w}\right)  $ by $w$ (or $\bar{w}$ ) is defined
by the following procedure. Move $w$ (or $\bar{w}$) to the right for each term
in $f\left(  w,\bar{w}\right)  $ with the help of (\ref{N-C0}) (\ref{N-C}),
and then differentiate by $w$ (or $\bar{w}$) on the right-hand side. This
definition of the differentiation can also be described by use of the
commutator as \
\begin{equation}
\partial_{w}=-\theta^{-1}\left[  \bar{w},\;\right]  \sigma_{3},\;\partial
_{\bar{w}}=\theta^{-1}\left[  w,\;\right]  \sigma_{3}\;. \label{z2deff}%
\end{equation}
Because of the nilpotency of $w$ and $\bar{w}$ (\ref{N-C0}), arbitrary
function of $Z_{2}$ space can be expanded in five terms,%
\begin{equation}
1,\;w\;,\;\bar{w},\;w\bar{w},\text{ \ }\bar{w}w. \label{term}%
\end{equation}
Here, four terms are linearly independent under the relation (\ref{N-C}).
Explicit form of the differentials are given by%
\begin{align}
\partial_{w}1  &  =\partial_{w}\bar{w}=0,\;\partial_{w}w=1,\nonumber\\
\;\partial_{w}w\bar{w}  &  =-\bar{w},\;\partial_{w}\bar{w}w=\bar{w}%
\end{align}
and%
\begin{align}
\partial_{\bar{w}}1  &  =\partial_{\bar{w}}w=0,\;\partial_{\bar{w}}\bar
{w}=1,\nonumber\\
\;\partial_{\bar{w}}w\bar{w}  &  =w,\;\partial_{\bar{w}}\bar{w}w=-w.
\end{align}
These can also be represented by matrix form, corresponding to
(\ref{zzb-matrix}), as%
\begin{equation}
\partial_{w}\left(
\begin{array}
[c]{cc}%
A & B\\
C & D
\end{array}
\right)  =\frac{1}{\sqrt{\theta}}\left(
\begin{array}
[c]{cc}%
B & 0\\
-A+D & B
\end{array}
\right)  \label{diff-z}%
\end{equation}
and%
\begin{equation}
\partial_{\bar{w}}\left(
\begin{array}
[c]{cc}%
A & B\\
C & D
\end{array}
\right)  =\frac{1}{\sqrt{\theta}}\left(
\begin{array}
[c]{cc}%
C & A-D\\
0 & C
\end{array}
\right)  , \label{diff-zb}%
\end{equation}
where we used the fact that%
\begin{equation}
\left(
\begin{array}
[c]{cc}%
A & B\\
C & D
\end{array}
\right)  =A\frac{w\bar{w}}{\theta}+B\frac{w}{\sqrt{\theta}}+C\frac{\bar{w}%
}{\sqrt{\theta}}+D\frac{\bar{w}w}{\theta}. \label{matrix-rep}%
\end{equation}
Furthermore, the integral in $w$ space is defined by the super trace on the
Fock space $\left\{  \left\vert 0\right\rangle ,\left\vert 1\right\rangle
\right\}  $ as
\begin{equation}%
{\displaystyle\int\nolimits_{Z_{2}}}
Od^{2}w=\text{str}O\mathrm{=\theta}\left\{  \left\langle 0\right\vert
O\left\vert 0\right\rangle -\left\langle 1\right\vert O\left\vert
1\right\rangle \right\}  , \label{int}%
\end{equation}
because of the anti-commutation relations (\ref{N-C0}) and (\ref{N-C}). In the
following, we take $\theta=1$ for simplicity.

\section{Vortices in $R^{2}$ space as instantons in $R^{2}\times Z_{2}$ space}

In this section, we discuss the YMH model in $R^{2}$ space which descends from
the Yang-Mills (YM) model in $R^{2}\times Z_{2}$ space, where $Z_{2}$ is the
noncommutative discrete space. The following is the discussion on the
self-dual equations in 4-dimensional $R^{2}\times Z_{2}$ space and BPS
equations for the vortex in 2-dimensional $R^{2}$ space. First, we sketch the
argument in ref. \cite{Leznov:1992ue} on the self-dual equations in pure
$U(N)$ YM model in commutative $R^{4}$ space\textit{.} As we shall see later,
applying this discussion to the $R^{2}\times Z_{2}$ space, BPS equations in
YMH model can be obtained.

Their\textit{ }argument\ goes as follows. They consider the self-dual equation
for pure $U(N)$ YM model in commutative $R^{4}$ space. From the relation
(\ref{leznov1}), the self-dual equation
\begin{equation}
F_{\mu\nu}=\frac{1}{2}\epsilon_{\mu\nu\rho\sigma}F_{\rho\sigma} \label{sd-eq}%
\end{equation}
can be rewritten as%

\begin{align}
F_{\bar{z}w}  &  =0,\label{8-1}\\
\;F_{z\bar{w}}  &  =0,\label{8-1b}\\
F_{z\bar{z}}  &  =F_{w\bar{w}}. \label{8-2}%
\end{align}
in commutative $z$ and $w$ coordinates. In this model, two $U(N)$ matrix
functions $h$ and $\bar{h}$ are introduced with the definition of gauge field
as%
\begin{align}
a_{z}  &  =h^{-1}\partial_{z}h,\;a_{\bar{z}}=\bar{h}^{-1}\partial_{\bar{z}%
}\bar{h},\nonumber\\
a_{\bar{w}}  &  =h^{-1}\partial_{\bar{w}}h,\;a_{w}=\bar{h}^{-1}\partial
_{w}\bar{h}\;.
\end{align}
Then, a part of self-dual equations (\ref{8-1}) and (\ref{8-1b}) are satisfied
automatically. And from
\begin{equation}
F_{z\bar{z}}=h^{-1}\partial_{z}\left(  g^{-1}\partial_{\bar{z}}g\right)
h,\;F_{w\bar{w}}=-\left(  h^{-1}\partial_{w}\left(  g^{-1}\partial_{\bar{w}%
}g\right)  h\right)  ,
\end{equation}
where%
\begin{equation}
g=\bar{h}h^{-1},
\end{equation}
another equation (\ref{8-2}) takes the form%
\begin{equation}
\partial_{z}\left(  g^{-1}\partial_{\bar{z}}g\right)  +\partial_{w}\left(
g^{-1}\partial_{\bar{w}}g\right)  =0. \label{yang}%
\end{equation}
Equation (\ref{yang}) is called Yang equation \cite{Yang:1977zf}.

To apply the above argument to the case of $R^{2}\times Z_{2}$ space, where
$Z_{2}$ space is noncommutative defined by (\ref{z2}), (\ref{z22}), we have to
replace the coordinates $\left(  w,\bar{w}\right)  $ in the previous argument
by noncommutative discrete ones for the equations from (\ref{coordinate}) to
(\ref{leznov1}) and from (\ref{sd-eq}) to (\ref{yang}). Especially, the
self-dual equations are
\begin{align}
F_{\bar{z}w}  &  =0,\nonumber\\
\;F_{z\bar{w}}  &  =0,\nonumber\\
F_{z\bar{z}}  &  =F_{w\bar{w}}, \label{sd-eq-z2}%
\end{align}
where $\left(  z,\bar{z}\right)  $ are the commutative $R^{2}$\ coordinates
and $\left(  w,\bar{w}\right)  $ are the noncommutative $Z_{2}$\ ones.
Futhermore, $h$ and $\bar{h}$ are expressed by the $\left(  N\times N\right)
\otimes\left(  2\times2\right)  $ matrices
\begin{align}
h  &  =\left(
\begin{array}
[c]{cc}%
0 & b\\
c & 0
\end{array}
\right)  ,\;h^{-1}=\left(
\begin{array}
[c]{cc}%
0 & c^{-1}\\
b^{-1} & 0
\end{array}
\right)  ,\label{h1}\\
\bar{h}  &  =\left(
\begin{array}
[c]{cc}%
0 & \bar{b}\\
\bar{c} & 0
\end{array}
\right)  ,\;\bar{h}^{-1}=\left(
\begin{array}
[c]{cc}%
0 & \bar{c}^{-1}\\
\bar{b}^{-1} & 0
\end{array}
\right)  . \label{h2}%
\end{align}
Namely, $h$ and $\bar{h}$ are expressed as $2\times2$ matrices (\ref{h1}),
(\ref{h2}), and each matrix elements $b,c,\bar{b},\bar{c}$ are $U(N)$
matrices. Replacing $R^{4}$ space by $R^{2}\times Z_{2}$ space, the gauge
field corresponding to $Z_{2}$ space can be considered as the Higgs field. In
the following, we show the equivalence between the instanton of YM model in
$R^{2}\times Z_{2}$ space and the vortex of YMH model in$\ R^{2}$ space.

Now, we shall consider the gauge fields and strengths. We use the differential
rules (\ref{z2deff}) or (\ref{diff-z}) (\ref{diff-zb}) for the $Z_{2}%
$\ coordinates. The gauge fields are given by%
\begin{align}
a_{z}  &  =h^{-1}\partial_{z}h=\left(
\begin{array}
[c]{cc}%
c^{-1}\partial_{z}c & 0\\
0 & b^{-1}\partial_{z}b
\end{array}
\right)  ,\label{gauge-a}\\
\;a_{\bar{z}}  &  =\bar{h}^{-1}\partial_{\bar{z}}\bar{h}=\left(
\begin{array}
[c]{cc}%
\bar{c}^{-1}\partial_{\bar{z}}\bar{c} & 0\\
0 & \bar{b}^{-1}\partial_{\bar{z}}\bar{b}%
\end{array}
\right)  , \label{gauge-ab}%
\end{align}%
\begin{align}
a_{w}  &  =\bar{h}^{-1}\partial_{w}\bar{h}\nonumber\\
&  =\left(
\begin{array}
[c]{cc}%
0 & \bar{c}^{-1}\\
\bar{b}^{-1} & 0
\end{array}
\right)  \partial_{w}\left(
\begin{array}
[c]{cc}%
0 & \bar{b}\\
\bar{c} & 0
\end{array}
\right) \nonumber\\
&  =\left(
\begin{array}
[c]{cc}%
0 & \bar{c}^{-1}\bar{b}\\
1 & 0
\end{array}
\right)  \label{gauge-z2}%
\end{align}
and%
\begin{align}
a_{\bar{w}}  &  =h^{-1}\partial_{\bar{w}}h\nonumber\\
&  =\left(
\begin{array}
[c]{cc}%
0 & c^{-1}\\
b^{-1} & 0
\end{array}
\right)  \partial_{\bar{w}}\left(
\begin{array}
[c]{cc}%
0 & b\\
c & 0
\end{array}
\right) \nonumber\\
&  =\left(
\begin{array}
[c]{cc}%
0 & 1\\
b^{-1}c & 0
\end{array}
\right)  . \label{gauge--z2b}%
\end{align}
Then we define the gauge fields $L$, $R$ and the Higgs field $H$\ as%
\begin{align}
L_{z}  &  =c^{-1}\partial_{z}c,\;L_{\bar{z}}=\bar{c}^{-1}\partial_{\bar{z}%
}\bar{c},\nonumber\\
R_{z}  &  =b^{-1}\partial_{z}b,\;R_{\bar{z}}=\bar{b}^{-1}\partial_{\bar{z}%
}\bar{b} \label{gauge--f}%
\end{align}
and%
\begin{equation}
H=\bar{c}^{-1}\bar{b},\;H^{\dagger}=b^{-1}c, \label{higgs-f}%
\end{equation}
respectively. Here, $h$ and $\bar{h}$\ are related as%
\begin{equation}
h^{\dagger}=\bar{h}^{-1}\text{ or\ \ \ ( }b^{\dagger}=\bar{b}^{-1},\text{
}c^{\dagger}=\bar{c}^{-1}\text{\ ),}%
\end{equation}
and the gauge fields are anti-Hermitan%
\begin{equation}
L_{z}^{\dagger}=-L_{\bar{z}},\;R_{z}^{\dagger}=-R_{\bar{z}}.
\end{equation}
The field strengths are calculated as follows. First, $F_{\bar{z}w}$ and
$F_{z\bar{w}}$ are calculated as
\begin{align}
F_{\bar{z}w}  &  =\partial_{\bar{z}}a_{w}-\partial_{w}a_{\bar{z}}+\left[
a_{\bar{z}},a_{w}\right] \nonumber\\
&  =\partial_{\bar{z}}\left(
\begin{array}
[c]{cc}%
0 & \bar{c}^{-1}\bar{b}\\
1 & 0
\end{array}
\right)  -\partial_{w}\left(
\begin{array}
[c]{cc}%
\bar{c}^{-1}\partial_{\bar{z}}\bar{c} & 0\\
0 & \bar{b}^{-1}\partial_{\bar{z}}\bar{b}%
\end{array}
\right) \nonumber\\
&  +\left[  \left(
\begin{array}
[c]{cc}%
\bar{c}^{-1}\partial_{\bar{z}}\bar{c} & 0\\
0 & \bar{b}^{-1}\partial_{\bar{z}}\bar{b}%
\end{array}
\right)  ,\left(
\begin{array}
[c]{cc}%
0 & \bar{c}^{-1}\bar{b}\\
1 & 0
\end{array}
\right)  \right] \nonumber\\
&  =\left(
\begin{array}
[c]{cc}%
0 & \partial_{\bar{z}}\left(  \bar{c}^{-1}\bar{b}\right)  +\left(  \bar
{c}^{-1}\partial_{\bar{z}}\bar{c}\right)  \bar{c}^{-1}\bar{b}-\bar{c}^{-1}%
\bar{b}\left(  \bar{b}^{-1}\partial_{\bar{z}}\bar{b}\right) \\
0 & 0
\end{array}
\right) \nonumber\\
&  =\left(
\begin{array}
[c]{cc}%
0 & D_{\bar{z}}H\\
0 & 0
\end{array}
\right)  \label{fs-zz2-2}%
\end{align}
and%
\begin{align}
F_{z\bar{w}}  &  =\partial_{z}a_{\bar{w}}-\partial_{\bar{w}}a_{z}+\left[
a_{z},a_{\bar{w}}\right] \nonumber\\
&  =\partial_{z}\left(
\begin{array}
[c]{cc}%
0 & 1\\
b^{-1}c & 0
\end{array}
\right)  -\partial_{\bar{w}}\left(
\begin{array}
[c]{cc}%
c^{-1}\partial_{z}c & 0\\
0 & b^{-1}\partial_{z}b
\end{array}
\right) \nonumber\\
&  +\left[  \left(
\begin{array}
[c]{cc}%
c^{-1}\partial_{z}c & 0\\
0 & b^{-1}\partial_{z}b
\end{array}
\right)  ,\left(
\begin{array}
[c]{cc}%
0 & 1\\
b^{-1}c & 0
\end{array}
\right)  \right] \nonumber\\
&  =\left(
\begin{array}
[c]{cc}%
0 & 0\\
\partial_{z}\left(  b^{-1}c\right)  +\left(  b^{-1}\partial_{z}b\right)
b^{-1}c-b^{-1}c\left(  c^{-1}\partial_{z}c\right)  & 0
\end{array}
\right) \nonumber\\
&  =\left(
\begin{array}
[c]{cc}%
0 & 0\\
D_{z}H^{\dagger} & 0
\end{array}
\right)  , \label{fs-zbzn-2}%
\end{align}
where%
\begin{equation}
D_{\bar{z}}H=\partial_{\bar{z}}H+L_{\bar{z}}H-HR_{\bar{z}}%
\end{equation}
and
\begin{equation}
D_{z}H^{\dagger}=\left(  D_{\bar{z}}H\right)  ^{\dagger}=\partial
_{z}H^{\dagger}-H^{\dagger}L_{z}+R_{z}H^{\dagger}.
\end{equation}
Note that the commutator term $\left[  a,a\right]  $ is needed even for the
$U(1)$ case because of the noncommutativity of $Z_{2}$ space. As in the case
of $R^{4}$,%

\begin{equation}
F_{\bar{z}w}=0 \label{f1}%
\end{equation}
and%
\begin{equation}
F_{z\bar{w}}=0 \label{f2}%
\end{equation}
are satisfied automatically with the definition of gauge fields by $h$ and
$\bar{h}$. Equations (\ref{fs-zz2-2}) (\ref{f1}) and (\ref{fs-zbzn-2})
(\ref{f2}) mean%
\begin{equation}
D_{\bar{z}}H=0 \label{bpss-1}%
\end{equation}
and
\begin{equation}
D_{z}H^{\dagger}=\left(  D_{\bar{z}}H\right)  ^{\dagger}=0 \label{bpssb-1}%
\end{equation}
respectively, and are nothing but a part of BPS equations for YMH model in
$R^{2}$.

Similarly,%

\begin{equation}
F_{zw}=\left(
\begin{array}
[c]{cc}%
0 & D_{z}H\\
0 & 0
\end{array}
\right)  ,\;F_{\bar{z}\bar{w}}=\left(
\begin{array}
[c]{cc}%
0 & 0\\
D_{\bar{z}}H^{\dagger} & 0
\end{array}
\right)
\end{equation}
are derived, where%
\begin{equation}
D_{z}H=\partial_{z}H+L_{z}H-HR_{z}%
\end{equation}
and%
\begin{equation}
D_{\bar{z}}H^{\dagger}=\left(  D_{z}H\right)  ^{\dagger}=\partial_{\bar{z}%
}H^{\dagger}-H^{\dagger}L_{\bar{z}}+R_{\bar{z}}H^{\dagger}.
\end{equation}

Finally, $F_{w\bar{w}}$ becomes%
\begin{align}
F_{w\bar{w}}  &  =\partial_{w}a_{\bar{w}}-\partial_{\bar{w}}a_{w}+\left[
a_{w},a_{\bar{w}}\right] \nonumber\\
&  =\partial_{w}\left(
\begin{array}
[c]{cc}%
0 & 1\\
b^{-1}c & 0
\end{array}
\right)  -\partial_{\bar{w}}\left(
\begin{array}
[c]{cc}%
0 & \bar{c}^{-1}\bar{b}\\
1 & 0
\end{array}
\right) \nonumber\\
&  +\left[  \left(
\begin{array}
[c]{cc}%
0 & \bar{c}^{-1}\bar{b}\\
1 & 0
\end{array}
\right)  ,\left(
\begin{array}
[c]{cc}%
0 & 1\\
b^{-1}c & 0
\end{array}
\right)  \right] \nonumber\\
&  =\left(
\begin{array}
[c]{cc}%
HH^{\dagger}-1 & 0\\
0 & -\left(  H^{\dagger}H-1\right)
\end{array}
\right)  , \label{fs-z2z2b}%
\end{align}
using \ (\ref{gauge-z2}) and (\ref{gauge--z2b}). $F_{z\bar{z}}$ is calculated
as%
\begin{align}
F_{z\bar{z}}  &  =\partial_{z}a_{\bar{z}}-\partial_{\bar{z}}a_{z}+\left[
a_{z},a_{\bar{z}}\right] \nonumber\\
&  =\partial_{z}\left(
\begin{array}
[c]{cc}%
\bar{c}^{-1}\partial_{\bar{z}}\bar{c} & 0\\
0 & \bar{b}^{-1}\partial_{\bar{z}}\bar{b}%
\end{array}
\right)  -\partial_{\bar{z}}\left(
\begin{array}
[c]{cc}%
c^{-1}\partial_{z}c & 0\\
0 & b^{-1}\partial_{z}b
\end{array}
\right) \nonumber\\
&  +\left(
\begin{array}
[c]{cc}%
\left[  c^{-1}\partial_{z}c,\bar{c}^{-1}\partial_{\bar{z}}\bar{c}\right]  &
0\\
0 & \left[  b^{-1}\partial_{z}b,\bar{b}^{-1}\partial_{\bar{z}}\bar{b}\right]
\end{array}
\right) \nonumber\\
&  =-i\left(
\begin{array}
[c]{cc}%
F_{12}^{L} & 0\\
0 & F_{12}^{R}%
\end{array}
\right)  . \label{fs-zzb}%
\end{align}
From (\ref{fs-z2z2b}) and (\ref{fs-zzb}), the self-dual equation (\ref{8-2})%
\begin{equation}
F_{z\bar{z}}=F_{w\bar{w}}%
\end{equation}
reduces to the BPS equations%
\begin{align}
iF_{12}^{L}  &  =1-HH^{\dagger},\\
iF_{12}^{R}  &  =H^{\dagger}H-1.
\end{align}
These are also expressed by Yang equation%
\begin{equation}
\partial_{z}\left(  g^{-1}\partial_{\bar{z}}g\right)  +\partial_{w}\left(
g^{-1}\partial_{\bar{w}}g\right)  =0,
\end{equation}
where%
\begin{equation}
g=\bar{h}h^{-1},
\end{equation}
and $h,\;\bar{h}$ are given by (\ref{h1}) and (\ref{h2}).

The above argument shows that the vortex in $R^{2}$ space can be regarded as
an instanton in $R^{2}\times Z_{2}$ space, since the self-dual equations
(\ref{sd-eq-z2}) of YM model in $R^{2}\times Z_{2}$ space is equivalent to the
BPS equations of YMH model in $R^{2}$ space.

Furthermore, we can see that the YM model in $R^{2}\times Z_{2}$ space also
reduces to the YMH model in $R^{2}$ space at the level of static part of the
Lagrangian. For the static configurations, square of field strength becomes%

\begin{align}
\frac{1}{4}\left\vert F_{^{\mu\upsilon}}\right\vert ^{2}  &  =\frac{1}%
{2}\left\vert F_{z\bar{z}}\right\vert ^{2}+\frac{1}{2}\left\vert F_{w\bar{w}%
}\right\vert ^{2}+\frac{1}{2}F_{zw}F_{\bar{z}\bar{w}}+\frac{1}{2}F_{\bar{z}%
w}F_{z\bar{w}{}}\nonumber\\
&  \equiv\left(
\begin{array}
[c]{cc}%
L_{1} & 0\\
0 & L_{2}%
\end{array}
\right)  , \label{Y-M-lag1}%
\end{align}
where%
\begin{align}
L_{1}  &  =\frac{1}{2}\left(  F_{12}^{L}\right)  ^{2}+\frac{1}{2}D_{\bar{z}%
}HD_{z}H^{\dagger}+\frac{1}{2}\left(  1-HH^{\dagger}\right)  ^{2},\nonumber\\
L_{2}  &  =\frac{1}{2}\left(  F_{12}^{R}\right)  ^{2}+\frac{1}{2}D_{\bar{z}%
}H^{\dagger}D_{z}H+\frac{1}{2}\left(  1-H^{\dagger}H\right)  ^{2}.
\label{Y-M-lag2}%
\end{align}
Then, in the case of YM model for the $U(N)\times U(N)$ gauge fields and
1-Higgs field, the Lagrangian density of YM model in $R^{2}\times Z_{2}$ space
is given by
\begin{equation}
\mathcal{L}=\left(
\begin{array}
[c]{cc}%
\mathbf{Tr}L_{1} & 0\\
0 & \mathbf{Tr}L_{2}%
\end{array}
\right)  \sigma_{3},
\end{equation}
where $\mathbf{Tr}$ means the trace of $U(N)$ matrix and $\sigma_{3}$ comes
from the volume element derived from the metric of the $Z_{2}$ space. Then the
action $S$ is obtained as
\begin{align}
S  &  =\int_{R^{2}\times Z_{2}}\text{str}\mathcal{L}d^{2}xd^{2}w\nonumber\\
&  =\int\text{str}\mathcal{L}d^{2}x\nonumber\\
&  =\int\text{str}\left\{  \left(
\begin{array}
[c]{cc}%
\mathbf{Tr}L_{1} & 0\\
0 & \mathbf{Tr}L_{2}%
\end{array}
\right)  \sigma_{3}\right\}  d^{2}x\nonumber\\
&  =\int\mathbf{Tr}\left\{  \frac{1}{2}\left(  F_{12}^{L}\right)  ^{2}%
+\frac{1}{2}\left(  F_{12}^{R}\right)  ^{2}+\frac{1}{2}D_{z}HD_{\bar{z}%
}H^{\dagger}+\frac{1}{2}D_{\bar{z}}HD_{z}H^{\dagger}+\left(  1-HH^{\dagger
}\right)  ^{2}\right\}  d^{2}x. \label{LR2}%
\end{align}
This gives the action of the YMH model in $R^{2}$ space (\ref{YMH-L}) with
$g^{2}=2$ and $c=1$ for the static configurations. It can also be verified
that the instanton number, denoted as $Q_{I}$,\ in $R^{2}\times Z_{2}$ space
is just the vortex number in $R^{2}$ space as follows.%
\begin{align}
Q_{I}  &  \equiv-\frac{1}{8\pi}\int\text{str}\left\{  \left(  \mathbf{Tr}%
F\tilde{F}\right)  \sigma_{3}\right\}  d^{2}x\nonumber\\
&  =-\frac{1}{8\pi}\int\text{str}\left\{  \left(  \mathbf{Tr}\frac{1}%
{2}\epsilon_{\mu\nu\alpha\beta}F_{\mu\nu}F_{\alpha\beta}\right)  \sigma
_{3}\right\}  d^{2}x\nonumber\\
&  =-\frac{1}{2\pi}\int\text{str}\left\{  \mathbf{Tr}\left(  -F_{z\bar{z}%
}F_{w\bar{w}}-F_{z\bar{w}}F_{\bar{z}w}+F_{zw}F_{\bar{z}\bar{w}}\right)
\sigma_{3}\right\}  d^{2}x\nonumber\\
&  =-\frac{1}{2\pi}\int\left[  {}\right.  \mathbf{Tr}\left\{  iF_{12}%
^{L}\left(  HH^{\dagger}-1\right)  +iF_{12}^{R}\left(  1-H^{\dagger}H\right)
\right\} \nonumber\\
&  \mathbf{\ \ \ \ \ \ \ \ \ \ \ \ \ \ \ }+\mathbf{Tr}\left\{  -D_{\bar{z}%
}HD_{z}H^{\dagger}+D_{z}HD_{\bar{z}}H^{\dagger}\right\}  \left.  {}\right]
d^{2}x\nonumber\\
&  =\frac{i}{2\pi}\int\mathbf{Tr}\left(  F_{12}^{L}-F_{12}^{R}\right)
d^{2}x\nonumber\\
&  =Q_{L-R}. \label{ins-vortex}%
\end{align}

\section{BPS, master and half-ADHM equations}

In the first part of this section, we study the BPS equation, master equation
and half-ADHM equation for YMH models with $U(N)$ and $U(N)\times U(N)$ gauge
groups. In the latter part of this section, we study the relation between
formulations for soliton equation discussed in section 4 and that of master
equation plus half-ADHM equation. We show that, for these two models, the BPS
equation for the vortex with certain topological number can be expressed by
the master equation plus half-ADHM equation. Furthermore, we see that the
vortex solution in two models satisfies the common half-ADHM equation. In
addition, we comment on some Abelian and non-Abelian vortices in both YMH
models. Finally, we obtain the relation between the variables in the two formulations.

First, we summarize the YMH model with $U(N)$ gauge group \cite{Eto}. The
Lagrangian is
\begin{equation}
{\mathcal{L}}=\mathrm{\mathbf{Tr}}\left(  \frac{1}{2g^{2}}\left(
F^{L}\right)  ^{\mu\nu}\left(  F^{L}\right)  _{\mu\nu}+\left(  D_{\mu
}H\right)  ^{\dag}D^{\mu}H-\frac{g^{2}}{4}(HH^{\dag}-c\boldsymbol{1}_{N}%
)^{2}\,\right)  . \label{YMH-L1}%
\end{equation}
Where, we define a covariant derivative $D_{\mu}$ and field strength
$F_{\mu\nu}^{L}$ as%
\begin{equation}
D_{\mu}H=\partial_{\mu}H+L_{\mu}H\;
\end{equation}
and%
\begin{equation}
F_{\mu\nu}^{L}=\partial_{\mu}L_{\nu}-\partial_{\nu}L_{\mu}+\left[  L_{\mu
},L_{\nu}\right]  \ .
\end{equation}
We take $g^{2}=2$ and $c=1$ in the following. BPS equations are
\begin{equation}
D_{\bar{z}}H=\partial_{\bar{z}}H+L_{\bar{z}}H=0 \label{bps-11}%
\end{equation}
and%
\begin{equation}
iF_{12}=1-HH^{\dagger}. \label{bps-21}%
\end{equation}
Let us introduce a $N\times N$ invertible matrix $S\left(  z,\bar{z}\right)
\in GL\left(  N,\mathbb{C}\right)  $ and consider a gauge invariant quantity
defined by%
\begin{equation}
\Omega\left(  z,\bar{z}\right)  \equiv S\left(  z,\bar{z}\right)  S^{\dag
}\left(  z,\bar{z}\right)  .
\end{equation}
Then the Higgs field and gauge field can be written as
\begin{align}
H  &  =S^{-1}H_{0},\label{Higgs-sakai}\\
L_{\bar{z}}  &  =S^{-1}\partial_{\bar{z}}S. \label{gauge-sakai}%
\end{align}
Here, $H_{0}\left(  z\right)  $ is the $N\times N$ matrix and has elements
consisting of holomorphic functions of $z$. The first BPS equation
(\ref{bps-11}) could be solved for arbitrary $S$ on account of these
relations. And the second BPS equation (\ref{bps-21}) is written in the form
of%
\begin{equation}
\partial_{z}\left(  \Omega^{-1}\partial_{\bar{z}}\Omega\right)  =1-\Omega
^{-1}H_{0}H_{0}^{\dagger}. \label{master}%
\end{equation}
This equation is called master equation \cite{Eto} for the vortices. The
vortex number is given by%
\begin{equation}
Q\equiv\dfrac{i}{2\pi}\int dx_{1}dx_{2}\mathrm{\mathbf{Tr}}F_{12}=0,\pm
1,\pm2,\cdots\ . \label{vortex-n}%
\end{equation}
From the master equation, at $\left\vert z\right\vert \rightarrow\infty$
\begin{equation}
\Omega=H_{0}H_{0}^{\dag}%
\end{equation}
for vortex configurations, since the left side of (\ref{master}) is
\begin{equation}
\partial_{z}\left(  \Omega^{-1}\partial_{\bar{z}}\Omega\right)  =\left(
S^{\dagger}\right)  ^{-1}\left(  F_{12}^{L}\right)  S^{\dagger}\rightarrow
0\text{ \ \ at \ }\left\vert z\right\vert \rightarrow\infty.
\end{equation}
\ Then, the vortex number (\ref{vortex-n}) can be rewritten as%
\begin{align}
Q  &  =k=\dfrac{1}{4\pi}\operatorname{Im}\oint dz\partial_{z}\log\left(  \det
H_{0}H_{0}^{\dagger}\right)  \ \nonumber\\
&  =\dfrac{1}{2\pi}\operatorname{Im}\oint dz\partial_{z}\log\left(  \det
H_{0}\right)  . \label{vortex number}%
\end{align}
This representation for the topological charge makes it clear that $H_{0}$
behaves like $\det H_{0}\sim z^{k}$ \ at the spacial infinity $\left\vert
z\right\vert \rightarrow\infty$. Moreover, $H_{0}(z)$ can be considered as a
solution of the half-ADHM equation \cite{Eto}%
\begin{equation}
\nabla^{\dagger}L=0. \label{h-adhm}%
\end{equation}
Here,
\begin{align}
L^{\dagger}  &  \equiv\left(  H_{0}\left(  z\right)  ,J\left(  z\right)
\right)  ,\nonumber\\
\nabla &  \equiv\left(
\begin{array}
[c]{c}%
-\Psi\\
z-Z
\end{array}
\right)  ,
\end{align}
and $H_{0}$, $J$, $\Psi$ and $Z$ are $N\times N$, $k\times N$, $k\times N$ and
$k\times k$ matrices, respectively. $\Psi$ and $Z$ are constant matrices and
have a meaning of moduli parameters. As a result, BPS equations reduce to the
master equation plus half-ADHM equation by introducing variables $S$ and
$H_{0}$. Here, $H_{0}$ is given as a solution of the half-ADHM equation. And,
for given $H_{0}$, $S$ is solved as a solution of the master equation.

Next, we extend the above argument to the case of $U(N)\times U(N)$ gauge
fields ($L_{\mu}$ and $R_{\mu}$) (\ref{YMH-L}). The BPS equations are%
\begin{equation}
D_{\bar{z}}H=\partial_{\bar{z}}H+L_{\bar{z}}H-HR_{\bar{z}}=0, \label{bps-3}%
\end{equation}%
\begin{align}
iF_{12}^{L}  &  =1-HH^{\dagger},\label{bps-4}\\
iF_{12}^{R}  &  =H^{\dagger}H-1. \label{bps-5}%
\end{align}
Expressing the Higgs field and gauge field as%
\begin{equation}
H=S^{-1}\left(  z,\bar{z}\right)  H_{0}\left(  z\right)  T\left(  z,\bar
{z}\right)  , \label{higgs-cf}%
\end{equation}%
\begin{align}
L_{\bar{z}}  &  =S^{-1}\partial_{\bar{z}}S,\\
R_{\bar{z}}  &  =T^{-1}\partial_{\bar{z}}T,
\end{align}
BPS equation (\ref{bps-3}) is satisfied automatically and BPS equations
(\ref{bps-4}) (\ref{bps-5}) are reduced to the two master equations
\begin{align}
\partial_{z}\left(  \Omega_{S}^{-1}\partial_{\bar{z}}\Omega_{S}\right)   &
=1-\Omega_{S}^{-1}H_{0}\Omega_{T}H_{0}^{\dagger},\nonumber\\
\partial_{z}\left(  \Omega_{T}^{-1}\partial_{\bar{z}}\Omega_{T}\right)   &
=-1+H_{0}^{\dagger}\Omega_{S}^{-1}H_{0}\Omega_{T}, \label{master-2}%
\end{align}
where
\begin{align}
\Omega_{S}  &  \equiv S\left(  z,\bar{z}\right)  S\left(  z,\bar{z}\right)
^{\dagger},\nonumber\\
\Omega_{T}  &  \equiv T\left(  z,\bar{z}\right)  T\left(  z,\bar{z}\right)
^{\dagger}.
\end{align}
\ 

At $\left\vert z\right\vert \rightarrow\infty$, we can see the following.
Finite energy of the static energy (\ref{YMH-E}) means that $U(N)\times U(N)$
gauge fields $L_{\mu}$ and $R_{\mu}$ go to pure gauge configurations. It is
possible to send the $SU(N)\times SU(N)$ part of gauge fields to zero, because
of the homotopy
\begin{equation}
\pi_{1}\left(  SU\left(  N\right)  \right)  =0.
\end{equation}
Then $S$ and $T$ can be expressed by elements of $U(1)$ as
\begin{equation}
S\left(  z,\bar{z}\right)  =s\left(  z,\bar{z}\right)  \cdot\mathbf{1}%
_{N},\;T\left(  z,\bar{z}\right)  =t\left(  z,\bar{z}\right)  \cdot
\mathbf{1}_{N}, \label{omegaST}%
\end{equation}
where $s(z,\bar{z})$ and $t(z,\bar{z})$ are scalar functions. Defining
\begin{equation}
S^{\prime}\equiv s\left(  z,\bar{z}\right)  t\left(  z,\bar{z}\right)  ^{-1},
\end{equation}
Higgs field (\ref{higgs-cf}) and $U(1)$ part of the gauge field are expressed
as%
\begin{equation}
H=S^{\prime}H_{0}%
\end{equation}
and%
\begin{equation}
\mathrm{\mathbf{Tr}}\left(  L_{\bar{z}}-R_{\bar{z}}\right)  =\left(
S^{\prime}\right)  ^{-1}\partial_{\bar{z}}S^{\prime}.
\end{equation}
Then, by the replacement $S\rightarrow S^{\prime}$, the topological charge
(\ref{vortex-n}) in $U(N)$ YMH model reduces to that in $U(N)\times U(N)$ model.

As a result, vortex number $Q_{L-R}$, given by (\ref{Qv}), can be expressed by%
\begin{equation}
Q_{L-R}=\dfrac{1}{4\pi}\operatorname{Im}\oint dz\partial_{z}\log\left(  \det
H_{0}H_{0}^{\dagger}\right)  ,
\end{equation}
which is same as (\ref{vortex number}). Therefore, $H_{0}$ satisfies the
common half-ADHM equation in each case of YMH model with $U(N)$ and $U\left(
N\right)  \times U(N)$ gauge groups. On the other hand, the master equation
turns to the coupled equations for $\Omega_{S}$ and $\Omega_{T}$ in the YMH
model with $U\left(  N\right)  \times U(N)$ gauge group.

Here, we comment on the vortex solutions of $U(1)\times U(1)$ and $U(N)\times
U(N)$ YMH models. It is known that when $F_{12}^{\ast}$ and $H^{\ast}$\ are a
numerical vortex solution of $U(1)$ YMH model, a vortex solution of $U(N)$ YMH
model can be constructed by embedding this vortex solution as%
\begin{equation}
F_{12}=U\mathrm{diag}(F_{12}^{\ast},0,\cdot\cdot\cdot,0)U^{-1}%
,\;H=U\mathrm{diag}(H^{\ast},1,\cdot\cdot\cdot,1)U^{-1},
\end{equation}
where $U$ takes a value in $CP^{N-1}$ \cite{Eto}. \ On the other hand, we can
show that a vortex solution of $U(1)\times U(1)$ YMH model is expressed by
that of $U(1)$ model by comparing both BPS equations. That is, denoting a
vortex configuration with topological number $m$\ of $U(1)$ model
(\ref{YMH-L1}) with $g^{2}=4$ and $c=1$ as $\tilde{F}_{12}^{\ast}$ and
$\tilde{H}^{\ast}$, a vortex of the $U(1)\times U(1)$ YMH model (\ref{YMH-L})
(with $g^{2}=2$ and $c=1$) is given by%
\begin{align}
F_{12}^{L}  &  =-F_{12}^{R}=\frac{1}{2}\tilde{F}_{12}^{\ast},\nonumber\\
H  &  =\tilde{H}^{\ast}. \label{abelian-vortex}%
\end{align}
And a non-Abelian vortex of the $U(N)\times U(N)$ YMH model is constructed as%
\begin{align}
F_{12}^{L}  &  =U\mathrm{diag}(\frac{1}{2}\tilde{F}_{12}^{\ast},0,\cdot
\cdot\cdot,0)U^{-1},\;\nonumber\\
F_{12}^{R}  &  =U\mathrm{diag}(-\frac{1}{2}\tilde{F}_{12}^{\ast},0,\cdot
\cdot\cdot,0)U^{-1},\;\nonumber\\
H  &  =U\mathrm{diag}(\tilde{H}^{\ast},1,\cdot\cdot\cdot,1)U^{-1}.
\label{nonabelian-vortex}%
\end{align}
As mentioned in section 2, it is obvious that the topological charge
$Q_{L+R}=0$ for the Abelian vortex (\ref{abelian-vortex}) and non-Abelian
vortex (\ref{nonabelian-vortex}), since $\mathbf{Tr}\left(  F_{12}^{L}%
+F_{12}^{R}\right)  =0$. And charge%
\[
Q_{L-R}\equiv\dfrac{i}{2\pi}\int dx_{1}dx_{2}\mathbf{Tr}\left(  F_{12}%
^{L}-F_{12}^{R}\right)  =\dfrac{i}{2\pi}\int dx_{1}dx_{2}\tilde{F}_{12}^{\ast
}=m
\]
\ counts the vortex number.

Finally, we consider the relation between variables ($h$ and $\bar{h}$) and
variables ($S,\;T$ and $H_{0}$) in YMH model with $U\left(  N\right)  \times
U(N)$ gauge group. A relation for the variables can take the form
\begin{equation}
\bar{h}=\left(
\begin{array}
[c]{cc}%
0 & \bar{b}\\
\bar{c} & 0
\end{array}
\right)  =\left(
\begin{array}
[c]{cc}%
0 & H_{0}^{T}\left(  z\right)  T\\
H_{0}^{S}\left(  z\right)  S & 0
\end{array}
\right)  . \label{hb}%
\end{equation}
We can check that two formulations lead the same Higgs field and gauge fields.
As the formulation given above in this section, taking the variables
$S,T,H_{0}$, Higgs field and gauge fields are given by%
\begin{equation}
H=S^{-1}\left(  z,\bar{z}\right)  H_{0}\left(  z\right)  T\left(  z,\bar
{z}\right)  ,
\end{equation}%
\begin{equation}
L_{\bar{z}}=S^{-1}\partial_{\bar{z}}S,\;R_{\bar{z}}=T^{-1}\partial_{\bar{z}}T,
\end{equation}
respectively. On the other hand, for the formulation discussed in section 4,
taking the variable $\bar{h}$ as (\ref{hb}), Higgs field and gauge fields are
given by%
\begin{equation}
H=\bar{c}^{-1}\bar{b}=S^{-1}(H_{0}^{S})^{-1}H_{0}^{T}T,
\end{equation}%
\begin{align}
L_{\bar{z}}  &  =\bar{c}^{-1}\partial_{\bar{z}}\bar{c}=S^{-1}\partial_{\bar
{z}}S,\\
\;R_{\bar{z}}  &  =\bar{b}^{-1}\partial_{\bar{z}}\bar{b}=T^{-1}\partial
_{\bar{z}}T.
\end{align}
Then, the condition that the two formulations give the same fields is%
\begin{equation}
(H_{0}^{S}\left(  z\right)  )^{-1}H_{0}^{T}\left(  z\right)  =H_{0}\left(
z\right)  .
\end{equation}
There exists some ambiguity in the relations of variables. A simple relation
is given by
\begin{equation}
\bar{h}=\left(
\begin{array}
[c]{cc}%
0 & \bar{b}\\
\bar{c} & 0
\end{array}
\right)  =\left(
\begin{array}
[c]{cc}%
0 & H_{0}\left(  z\right)  T\left(  z,\bar{z}\right) \\
S\left(  z,\bar{z}\right)  & 0
\end{array}
\right)  .
\end{equation}

\section{Summary and Discussion}

In this paper, we have defined the coordinates for noncommutative $Z_{2}%
$\ space and have investigated the relation between the instantons in
$R^{2}\times Z_{2}$\ space and the vortices in $R^{2}$ space. We have shown
that the vortices of YMH model in $R^{2}$ space can be regarded as the
instantons of YM model in $R^{2}\times Z_{2}$ space. The BPS equation for the
vortices can be considered as a self-dual Yang-Mills equation and is related
to the ADHM equation. We also have obtained the relations between the master
equation for the vortices and the Yang equation for the instantons.

It may be expected that the ADHM method can also be applied to the
construction of the vortex solutions. However, extension of ADHM equation into
the $R^{2}\times Z_{2}$ space is not straightforward. The reason can be traced
to noncommutativity of $\partial_{w}$ and $\partial_{\bar{w}}$ $\left(
\text{or }\partial_{3}\text{ and \ }\partial_{4}\right)  $. Writing the Dirac
operators as
\begin{align}
\boldsymbol{D}_{x}  &  \equiv e^{\mu}\otimes D_{\mu}=e^{\mu}\otimes\left(
\partial_{\mu}+A_{\mu}\right)  ,\nonumber\\
\boldsymbol{\bar{D}}_{x}  &  \equiv\bar{e}_{\mu}\otimes D_{\mu}%
=-\boldsymbol{D}_{x}^{\dagger},
\end{align}
where%
\begin{equation}
e_{\mu}=\left(  -i\sigma_{i},1\right)  ,\bar{e}_{\mu}=\left(  i\sigma
_{i},1\right)
\end{equation}
are quaternions. Square of Dirac operators are written as
\begin{equation}
\boldsymbol{\bar{D}}_{x}\boldsymbol{D}_{x}=1_{\left[  2\right]  }\otimes
D_{\mu}D_{\mu}+i\eta_{\mu\nu}^{i\left(  +\right)  }\sigma_{i}\otimes D_{\mu
}D_{\nu}, \label{ddb}%
\end{equation}
where%
\begin{equation}
\eta_{\mu\nu}^{i\left(  \pm\right)  }=\epsilon_{i\mu\nu4}\pm\delta_{i\mu
}\delta_{\nu4}\mp\delta_{i\nu}\delta_{\mu4}%
\end{equation}
are 't Hooft symbols. The last term of equation (\ref{ddb}) can be written for
$R^{4}$ space as
\begin{align}
i\eta_{\mu\nu}^{i\left(  +\right)  }\sigma_{i}\otimes D_{\mu}D_{\nu}  &
=i\sigma_{1}\otimes\left\{  F_{23}+F_{14}\right\}  +i\sigma_{2}\otimes\left\{
-F_{13}+F_{24}\right\} \nonumber\\
&  +i\sigma_{3}\otimes\left\{  F_{12}+F_{34}\right\}  ,
\end{align}
and the condition%
\begin{equation}
\left[  \boldsymbol{\bar{D}}_{x}\boldsymbol{D}_{x},\sigma_{i}\right]  =0
\label{cond-sd}%
\end{equation}
leads to the (anti-)self-dual equation
\begin{equation}
F_{\mu\nu}=-\frac{1}{2}\epsilon_{\mu\nu\alpha\beta}F_{\alpha\beta}\ .
\end{equation}
For $R^{2}\times Z_{2}$ space, however, we have
\begin{align}
i\eta_{\mu\nu}^{i\left(  +\right)  }\sigma_{i}\otimes D_{\mu}D_{\nu}  &
=i\sigma_{1}\otimes\left\{  F_{23}+F_{14}\right\}  +i\sigma_{2}\otimes\left\{
-F_{13}+F_{24}\right\} \nonumber\\
&  +i\sigma_{3}\otimes\left\{  F_{12}+F_{34}+\left[  \partial_{3},\partial
_{4}\right]  \right\}  ,
\end{align}
and because of noncommutativity of $Z_{2}$ space (\ref{cond-sd}) does not lead
to the self-duality equation and we have to find a different constraint.
Furthermore, unlike the case of noncommutative ADHM, $\left[  \partial
_{3},\partial_{4}\right]  $ is not a constant, thus we have to find a
different modification. Consequently, it is possible that ADHM equations are
not pure algebraic equations but include differential equations in $R^{2}$
space. And this could be related to the fact that it is impossible to obtain
the vortex solutions analytically.

We have compared our YMH model that contains two gauge fields with YMH model
with only one gauge field. In the latter model, we can rewrite the BPS
equations into the master equation plus half-ADHM equation. We can do the same
in the former model, the BPS equations also reduce to the master plus
half-ADHM equations and the half-ADHM equations in both models coincide
exactly with each other. Furthermore we have studied both Abelian and
non-Abelian vortices and the interrelations among them.

Although we have defined our $Z_{2}$ through equations (\ref{z2}),
(\ref{z22}), there exist other possibilities and they are probably worthwhile
to be considered. Furthermore, it has been proposed that there exists similar
relation in the case of the model on compact Riemann surface \cite{Popov}. It
would be interesting to examine the relations between our work and their
approach. \bigskip\bigskip

\noindent{\Large \textbf{Acknowledgments}}

We would like to thank Akihiro Nakayama for his support and hospitality. We
also thank Yoshimitsu Matsui for discussions.

\newpage

\providecommand{\href}[2]{#2}\begingroup\raggedright

\endgroup  
\end{document}